# Commute with Community: Enhancing Shared Travel through Social Networks

Siyuan Tian

Tongji University, College of Electronic and Information Engineering, 2053422@tongji.edu.cn

Renjie Dai

Tongji University, School of Software Engineering, dairenjie123@tongji.edu.cn

Junhao Wang

Tongji University, College of Electronic and Information Engineering, 2050653@tongji.edu.cn

ZhengXiao He*

Tongji University, College of Electronic and Information Engineering, 1950095@tongji.edu.cn

Shared mobility redefines urban transportation, offering economic and environmental benefits by reducing pollution and urban congestion. Howerver, in the post-pandemic era, the shared mobility sector is grappling with a crisis of trust, particularly concerning passenger hesitancy towards shared transportation options.To address those problems, in this paper we take sociel network into consideration and propose a novel carpooling matching framework based on graph neural network and reinforcement learning,increasing the carpooling rate to **48%**, and reducing the average delay time to **6.1 minutes** and the average detour distance to **2.8 km**. Furthermore, we introduce an innovative metric, termed 'tolerance' for mobility sheduling models to effectively quantify users' sensitivity to social distancing. We conduct a sensitivity analysis to demonstrate that our model offers a viable approach to amplify the benefits, delivering resilient strategies for the advancement and proliferation of shared mobility incentives.

## 1 INTRODUCTION

In the post-pandemic era, significant changes have occurred in the structure of transportation. Due to health concerns, public skepticism towards shared mobility has increased, leading to a rise in private car usage. This shift exacerbates issues of traffic congestion, carbon emissions, and travel costs. In the United States, traffic congestion incurs an annual cost of approximately $121 billion, equivalent to 1% of the Gross Domestic Product, encompassing wasted time and additional fuel consumption[1] . In mega-cities like Shanghai, there is an urgent need for low-carbon transportation solutions due to rising CO2 emissions [2]. Addressing traffic congestion and air pollution has become a pivotal issue, and shared mobility offers a sustainable alternative to traditional transportation methods by reducing pollution and urban congestion.

---

\* Corresponding authors.

Shared mobility redefines urban transportation, proving beneficial both economically and environmentally. Platforms such as Uber and Lyft have expanded globally, with millions of users and hundreds of thousands of drivers [3]. DiDi Chuxing's carpooling services significantly reduce carbon emissions, with millions of users utilizing the service daily [4]. To further optimize shared mobility, this study introduces an algorithm utilizing graph neural networks[5][6][7] and a reinforcement learning Policy-Network[8], assessing its impact on society, drivers, passengers, and platforms. And still we need to find out a useful method in Longitudinal Data Analysis.[9]. It aims to reduce travel distances, emissions, and energy consumption; improve working conditions for drivers; enhance passengers' carpooling experience; and boost the operational efficiency and revenue of platforms. This comprehensive approach may propel the adoption of shared mobility in the post-pandemic era, balancing individual preferences with broader societal benefits.

In summary, we make the following contributions:

(i) **Social Network Integration**: we pioneer the integration of social network factors, considering social connections and urban planning standards to identify carpooling opportunities. By constructing a complex network of passenger shared mobility and a bipartite graph of passenger-location relationships, we offer a novel approach to understanding and enhancing carpooling dynamics.

(ii) **Tolerance Function for Social Distancing Sensitivity:** we introduce a tolerance function to quantify users' sensitivity to social distancing, marking an innovative contribution that adds a new dimension to the optimization of shared mobility services, addressing evolving user preferences and safety concerns.

(iii) **Results**: our state-of-the-art framework, which synergizes graph neural networks and reinforcement learning and incorporates social distancing considerations, demonstrates promising results in quantitative comparisons. Specifically, it optimizes carpooling rates with a peak of 48% when minimizing vehicle count, effectively manages waiting times with an average delay reduced to 6.120 minutes, and strategically minimizes unnecessary travel, reducing the average detour distance to 2841.38 meters. This multifaceted approach not only advances the shared mobility domain but also provides a fresh perspective on passenger tolerance to waiting times, paving the way for future innovations in this rapidly evolving field.

## 2 RELATED WORK

In shared mobility optimization, current research centers on drivers, passengers, and platform-centric approaches, each aiming to refine carpooling dispatch methods and enhance system performance. Driver-focused studies aim to reduce driving distance and completion time.

Passenger perspective research concentrates on improving carpooling rates and participation willingness. Kameswaran et al. [10] emphasize social connections in carpooling, while Cheng et al. [11] integrate social networks into a utility model for better match-making. Mahajan et al.[12] propose a novel three-tier SSN item recommendation architecture (3T-IEC*) which takes geographical and social influence into consideration. Fu et al. [13] propose a social utility-based approach for route selection, and Quercia et al. [14] blend LBSNs with geographic data for network construction, enhancing route recommendations. What's more, Yang et al.[15] propose Graph Differential Equation to model the continuity of users' interest and combine sequence- and graph-based models to obtain a time-serial graph.

Platform-oriented studies focus on maximizing order completions and revenue. Kieiner et al.'s [16] dynamic carpooling system employs an auction mechanism for order assignment, optimizing pick-up and drop-off sequencing. Vazifeh et al.[17] introduce a traffic network algorithm to reduce vehicle usage, and Asghari et al. [18]

apply a branch-and-bound framework in APART for optimal driver allocation, driving revenue maximization. Mühle et al.[19] predicts ride pooling efficiency and required fleet size without simulations. Lotze C et al.[20] propose adaptive stop pooling may substantially reduce travel time fluctuations while even improving the average travel time of ride sharing services, especially for high demand. Lotze C et al.[21] propose a dimensionless parameter to estimate the sustainability of ridepooling by quantifying the load on a ridepooling service, relating characteristic timescales of demand and supply.

Critically, these approaches also consider the balance between system efficiency and passenger experience. Kleiner et al. [22] propose a balanced objective of maximizing carpooling value while accommodating detour-related delays, aligning cost efficiency with passenger convenience. Fèvre et al.[23] propose a preferential optimization for each individual in the shared system.However, this research primarily addresses long-term commuter carpooling rather than real-time ride-splitting, indicating a variance in application and target demographic.

## 3 PROPOSED METHOD

### 3.1 Building Passenger Shared Mobility Network Based on Graph Network Theory

In this study, we initially procure and map the geographical information involved in the dataset. By matching address data points with their latitude and longitude to the nearest map nodes, we plot the Origin-Destination (OD) lines of user travel and store travel data for constructing optimal carpooling routes.This geographical information matching process is depicted in Figure 1. Upon establishing a shared mobility network for users, each user's travel is regarded as a node within the network. We apply two constraints to identify potential carpooling opportunities and establish connections between nodes [24][25].

The social network constraint uses a three-kilometer radius to determine if users can form a social network based on proximity. The travel time constraint requires users' travel times to meet specified accessibility criteria for carpooling..

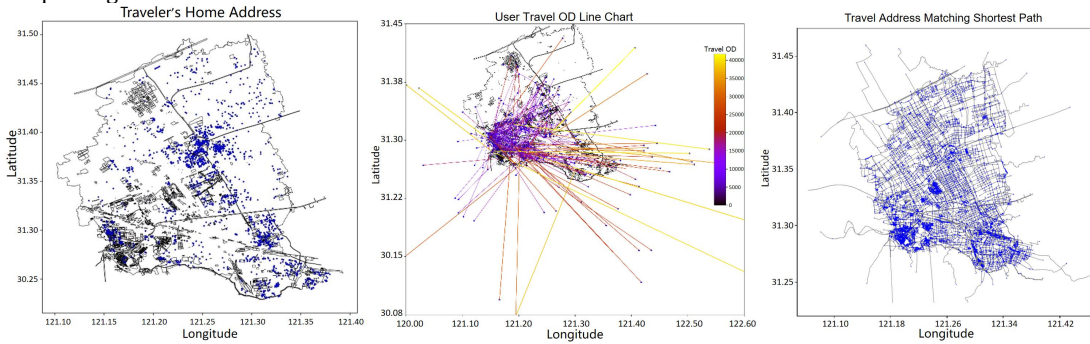

Figure 1: Geographic Information Matching Process

If two users fulfill both social network and travel time conditions, they can share a ride, forming a connection in the network graph. The edge weights, based on certain goals, help create the shared mobility network, shown in Figure 2, with specific weight assignments $W_{ij}$.

To maximize shared rides, weights are uniformly set at 2, where $W_{ij} = 2$ indicates a shared ride count of 2.

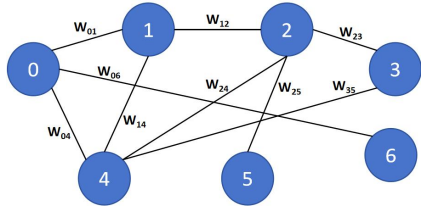

Figure 2: Passenger Shared Travel Network Concept Map

For the objective of minimizing the total travel distance of all trips, the weight $W_{ij}$ is defined as the reduction in travel distance due to ride-sharing.

$$W_{ij} = D_i^s + D_j^s - D_{ij}^t \qquad (1)$$

$D_i^s$ represents the distance traveled by user i when traveling alone. $D_{ij}^t$ denotes the total distance covered in a shared ride between users $i$ and $j$.

To achieve the objective of maximizing the total travel time of all trips, the weight $W_{ij}$ is defined as the reduction in travel time resulting from ride-sharing.

$$W_{ij} = T_i^s + T_j^s - T_{ij}^t \qquad (2)$$

$T_i^s$ represents the time traveled by user i when traveling alone. $T_{ij}^t$ denotes the total time covered in a shared ride between users $i$ and $j$.

### 3.2 Construction of User-Location Bipartite Graph for Shared Mobility

To further enhance our analysis, we aim to construct a bipartite graph that represents the relationship between users and locations. This graph is intended to capture geographical information about the places users frequently visit and leverage travel data to analyze user characteristics.

Let's assume we have two types of nodes: User $A$ and Location $B$. There is a connection between these two types of nodes, signifying that User $A$ has visited Location $B$.

We can describe this relationship using graph theory terminology: Let $G = (V, E)$ be a graph, where $V$ represents the set of nodes, and $E$ represents the set of edges. In this graph, $V$ consists of User $A$ and Location $B$, and $E$ includes pairs of (User $A$, Location $B$). To encode the entire dataset of locations, we employ a grid-based method based on latitude and longitude, connecting users to all the locations they have visited. This forms the user-location bipartite graph, as depicted in Figure 3 in the research paper.

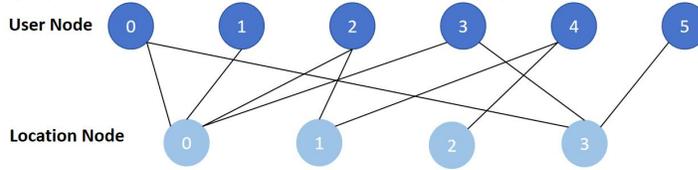

Figure 3: Passenger-Location Bipartite Graph

### 3.3 Calculating User Feature Vectors using the Neural Graph Collaborative Filtering Algorithm

After obtaining the user-location bipartite graph, we can calculate user feature vectors using the Neural Graph Collaborative Filtering algorithm. Neural Graph Collaborative Filtering is a graph neural network-based collaborative filtering algorithm, with its network structure illustrated in Figure 4. It combines the advantages of both collaborative filtering and neural networks, enabling more effective handling of sparse user-item interaction data.

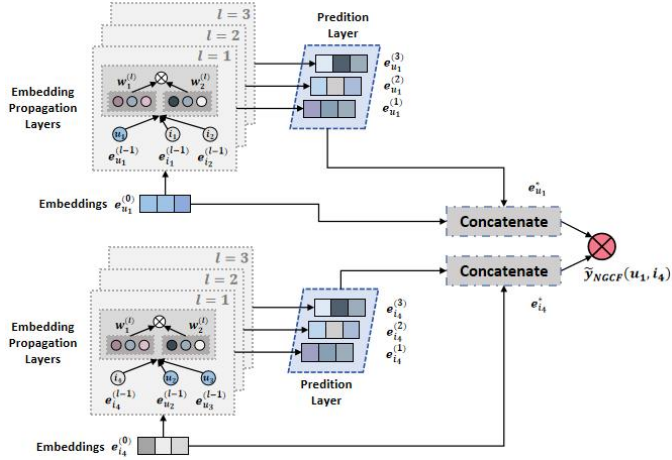

Figure 4: Neural Graph Collaborative Filtering Algorithm Schematic Diagram

The core idea of the Neural Graph Collaborative Filtering algorithm is to leverage a graph neural network to learn representations of users and locations from their interaction relationships. It achieves this by utilizing the bipartite graph of users and locations, propagating feature information through adjacency matrices, and employing connections and information aggregation across multiple layers of neural networks. This enables each node to perceive information from neighboring nodes and update its own representation. Consequently, the feature vectors of users and locations continuously adapt and improve as information propagates, thereby capturing the associations between users and locations.

In this research, the processing steps are as follows:

1. **Initialization of User and Location Feature Vectors**:

Let $U$ represent the set of users, and $W$ represent the set of locations. For each user in $U$ and location in $I$, initialize their feature vectors: $u$ represents the feature vector for user in $U$, and $i$ represents the feature vector for location in $I$.

2. **Definition of Adjacency Matrices**:

Based on the interaction matrix $A$, construct adjacency matrices for users and locations. Define two adjacency matrices, $E^U$ and $E^I$:

$E^U = A \cdot A^T$ represents the similarity between users (user-user adjacency matrix).

$E^I = A^T \cdot A$ represents the similarity between locations (location-location adjacency matrix).

3. **Feature Propagation**:

Utilize the user-location interaction matrix and adjacency matrices for feature propagation, updating the feature vectors of users and locations. The process of feature propagation can be represented as follows:

$$E^{(l)} = \sigma\left((L + I) \cdot E^{(l-1)} W_1^{(l)} + L E^{(l-1)} \odot E^{(l-1)} W_2^{(l)}\right) \quad (3)$$

$\sigma$ represents the activation function (such as ReLU or Sigmoid), $E^{(l)}$ denotes the vector representations of users and locations obtained after l iterations of embedding propagation, $I$ represents the identity matrix, and $L$ represents the graph Laplacian matrix.

4. **Repeat Feature Propagation Steps**:

Reiterate the feature propagation steps until convergence or a predetermined number of iterations is reached.

5. **Aggregate User Feature Vectors from Multiple Layers**:

After $L$ layers of propagation, we obtain multiple representations for user $u$ namely $e_u^{(1)}, e_u^{(2)}, ..., e_u^{(L)}$. Since these representations emphasize messages transmitted through different connections at different layers, they contribute differently to reflecting user preferences. Therefore, we concatenate them to form the final embedding for the user.

Once feature propagation converges, we obtain the updated user feature vector $e_u$, which can be utilized in subsequent shared mobility matching algorithms.

### 3.4 Building Shared Mobility Matching Solutions through Policy Network

Following the acquisition of user feature vectors, the problem of creating personalized carpooling strategies for specific users is formalized as a sequential decision-making process. Faced with the current environment comprising the user and already selected co-riders, the algorithm determines the addition of new carpooling users to the existing group. The reward function for each decision is designed to cater to varying user needs. The optimal shared mobility matching solution emerges from a sequence of decisions made by an intelligent agent. This problem is ultimately addressed using reinforcement learning methods, as illustrated in the reinforcement learning framework shown in Figure 5.

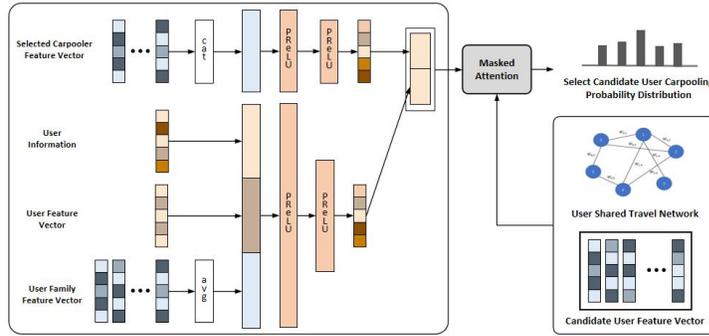

Figure 5: Visualization of Policy Network Obtaining Carpooler Probability Distribution

The formal definition of solving the shared mobility matching problem using reinforcement learning and a Policy Network is as follows:

• **Initial State**: The initial state is determined by the specific user attributes, namely the user context vector. The probability distribution for the initial state is determined based on the user context vector.

• **State Space**: The state space consists of three parts. The first part is the contextual vector $u$ derived from the Neural Graph Collaborative Filtering algorithm. The second part is the passenger shared mobility network $G$. The final part is the currently selected user feature sequence $V_t$. These three components together describe the state for a specific user at a specific moment, denoted as a triple $(u, G, V_t)$.

• **Action Space**: The action space involves selecting users from all potential candidates who are not currently part of the shared mobility set $V_t$ and adding them to the carpooling group.

• **State Transition**: Taking an action by selecting a node to be added to the previous state results in a new state. In the state transition, the user context vector and the passenger shared mobility network G remain unchanged, with only the addition of a new co-rider from the selected candidates. Therefore, the state after the state transition is:

$$s_{t+1} = (u, G, V_{t+1} = V_t \cup \{v\}) \qquad (4)$$

• **Reward Function**: Different reward functions can be used for different user preferences. For example, travel time, total distance, social distance, etc., can be chosen as reward functions.

Specifically, Proximal Policy Optimization algorithm proposed by OpenAI is utilized. It evaluates the policy for each action in each state based on the state space, parameters and trains the policy, and obtains the action probability distribution for each state. Sampling and selection of different actions from the returned action probability list are then performed to obtain the carpooling matching results[26].

## 4 RESULTS AND DISCUSSION

In evaluating the effectiveness of carpooling within the shared mobility framework, several key performance indicators were analyzed as demonstrated in Table 1. These indicators span a range of aspects from vehicle utilization to environmental impact, providing a comprehensive perspective on the practical application of carpooling algorithms.

Initially, the vehicle occupancy rate, a critical efficiency metric, is measured by the ratio of occupied kilometers to the total distance traveled. The findings indicated optimization of this ratio, reaching 1.531, when minimizing travel distance, thereby indicating enhanced vehicle utilization. Variations in the carpooling rate, which denotes the ratio of shared to total travels, were observed under different optimization strategies. A peak rate of 48% was achieved in reducing vehicle count, highlighting the capability to reduce active vehicles. Conversely, this objective led to the highest average delay time of 6.120 minutes, suggesting a need to balance between vehicle reduction and passenger convenience.

Assessment of average detour distance revealed a significant reduction to 2841.38 meters when minimizing vehicle count, indicating that strategic matching can reduce unnecessary travel. The consistency of detour ratios and discount rates across objectives illustrates the algorithm's ability to consistently reduce travel and save fares for passengers. Environmentally, the most significant emissions, calculated using the MOVES model, correlated with the objective of minimizing vehicle count, implying that fewer vehicles paradoxically might lead to increased emissions due to longer trips per vehicle. Finally, despite minimized detour distances, fuel costs still increased under reduced vehicle count, likely due to increased total vehicle usage, thereby elevating overall fuel expenditure.

These insights emphasize that different optimization objectives impact the operational outcomes of the carpooling system differently. Practical deployment necessitates aligning these objectives with the nuanced demands and policies of urban traffic governance to fully leverage the societal benefits of carpooling, such as reducing congestion, improving energy usage, minimizing pollution, and enhancing passenger satisfaction.

Table 1: Assessment of Ride-splitting Efficiency and Environmental Impact Indicators

| Evaluation indicators | Definition | O_distance | O_time | O_vehicle |
| --- | --- | --- | --- | --- |
| Vehicle Occupancy Rate | Passenger-mile efficiency | 1.531 | 1.341 | 1.227 |
| Carpooling Rate | Share of rides that are carpooled | 0.46 | 0.44 | 0.48 |
| Average Delay Time | Extra wait time for carpooled rides | 4.179 | 3.989 | 6.120 |
| Average Detour Distance | Additional travel distance for carpools | 4288.51 | 4302.76 | 2841.38 |
| Detour Ratio | Extra distance relative to direct distance | 0.2876 | 0.2765 | 0.1817 |
| Discount Ratio | Savings on fares for passengers | 0.2104 | 0.2096 | 0.2096 |
| Emissions | Calculated pollutants for the trips | 9874.9 | 10254.3 | 12435.2 |
| Fuel Consumption | Total fuel used for all trips | 52223.79 | 53224.27 | 53550.42 |

Integrating social distancing factors into the shared mobility framework provides a novel perspective for assessing passenger tolerance to waiting times. Analysis of data, pre- and post-integration of these factors as shown in Figure 6, indicates a significant shift in passenger sensitivity to waiting times.

Initial data, without considering social distancing, showed consistent tolerance levels across scenarios, indicating uniform passenger behavior. However, the inclusion of social distancing elements marks a significant change in waiting time sensitivity, reflecting the evolving preferences and tolerances of passengers. As social distancing becomes increasingly important, there is a noticeable decrease in tolerance to waiting, particularly in scenarios with fewer vehicles. This trend may be attributed to heightened preferences for personal space and safety, leading passengers to opt for private or less crowded transportation options, even at the cost of delays.

Conversely, sensitivities to reduced travel time and shorter distances remain relatively unaffected by social distancing considerations, suggesting a minimal impact on passengers' willingness to accept longer routes or minor service delays. These insights imply that in a post-pandemic context, where social distancing may continue to be a factor, passenger sensitivity to waiting times could become more critical, necessitating strategic shifts in optimization strategies for shared mobility platforms to align with new levels of passenger tolerance.

In summary, the incorporation of social distancing into shared mobility models reveals a complex interplay between passenger tolerance to waiting times and social distancing preferences, which is crucial for tailoring shared mobility services to meet the evolving expectations of passengers in the post-pandemic era.

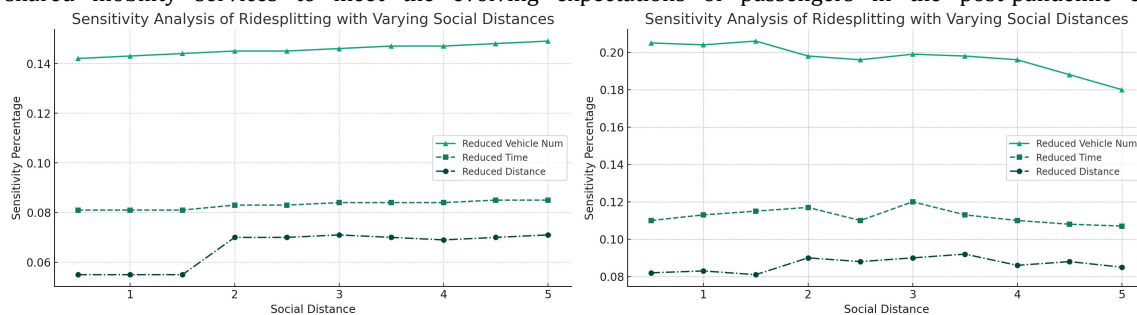

Figure 6: Sensitivity Analysis of Ride-splitting Performance Across Social Distances

## 5   CONCLUSIONS AND FUTURE RESEARCH

In the current backdrop of the pandemic, this research addresses the issue of trust crisis in the field of shared mobility by proposing and extensively analyzing an innovative social network-based carpooling model. By integrating graph theory and the Neural Graph Collaborative Filtering algorithm, a complex passenger shared mobility network is constructed, effectively capturing user characteristics and presenting a carpooling travel matching solution defined by a reward function. This solution is optimized using the Policy Network algorithm, successfully balancing key performance indicators such as time, distance, and cost.

In the process of model construction, the introduction of tolerance functions and heterogeneous parameters enhances the model's interpretative capabilities for individual differences, emphasizing the importance of considering user psychological needs in model design. Sensitivity analysis of these parameters reveals potential directions for optimizing model benefits, providing a theoretical basis for the incentive mechanism in shared mobility and promoting the practical application of shared mobility services.

Looking ahead, as social distancing becomes the new norm, this research model and its findings offer a fresh perspective on addressing user acceptance issues in shared mobility services. Future research can further explore the application of social network elements in other types of shared services, such as shared bicycles and electric scooters. Additionally, there is a need to study how to adapt and optimize the proposed model in different cultural and geographical contexts to meet diverse demands worldwide. Furthermore, with technological advancements and the accumulation of user behavioral data, there will be more opportunities to enhance the accuracy and scalability of the model using machine learning and artificial intelligence algorithms. Ultimately, these efforts will contribute to fostering societal acceptance of shared mobility and promoting its sustainable development on a global scale.